
\magnification=1200
\baselineskip 20pt

\vbox{\vskip 2cm}

\rightline{CINVESTAV-GR IPN 94/05}
\rightline{May, 1994}
\centerline {\bf A NOTE ON WEAK ${\cal H}$ - SPACES, HEAVENLY
EQUATIONS}
\centerline {\bf  AND THEIR EVOLUTION WEAK HEAVENLY
EQUATIONS}
\vskip.3cm
\centerline{ J.F. Pleba\'nski\footnote{\dag}{
On leave of absence from the University of Warsaw, Poland.}
 and H.
Garc\'\i a-Compe\'an\footnote{\ddag}{Supported by a CONACyT and SNI
Fellowships}}
\vskip.3cm
\centerline{{\it Departamento de F\'\i sica} }
\centerline{{\it Centro de Investigaci\'on y de
Estudios Avanzados del IPN.}}
\centerline{{\it Apdo. Postal 14-740, 07000, M\'exico D.F. M\'exico.}}
\vskip2cm

It is find a non-linear partial differential equation which we show
contains the first Heavenly equation of Self-dual gravity  and generalize
the second one. This differential equation we call "Weak Heavenly
Equation" (${\cal WH}$-equation). For the two-dimensional case the
${\cal WH}$-equation is brought
into the evolution form (Cauchy-Kovalevski form) using the  Legendre
transformation. Finally, we find that this transformed equation ("Evolution
Weak Heavenly Equation") does admit very simple solutions.

\vfill\eject
\centerline{ \bf 1. Introduction}

Complex Self-dual (SD) gravity has been a great arena to understand
new advances in other branches of mathematical physics; namely,
Quantum Gravity, Conformal Field Theory,
Integrable Models, Topological Field Theories, String Theory, etc. For this
reason, any advance in this
line is of great importance in the development of these branches.

Very recently has been a series of papers on SD gravity [1-5], based in the
seminal Grant's paper [6]. In they, the first and the second heavenly
equations and all hyper-heavenly equations with (and without) cosmological
constant were broughted into an evolution form called
Cauchy-Kovalevski form. Some solutions for all these evolution equations
also has been found. In all cases the relation between the heavenly
equations
and that of evolution has been a Legendre transformation applied on suitable
coordinates.

On the other hand, the richness of structure in SD gravity allow us
to explore what another possibilities might exists there. In this paper we
explore
one of these possibilities. We find that working with the spinorial
formulation of SD gravity we find a non-linear partial differential
equation of the second order. This equation we call "Weak Heavenly
equation" (${\cal WH}$-equation) and involves a holomorphic
function of its arguments. This equation contains the first Heavenly
equation  and in fact  generalize
the second one. Taking the dimensional reduction of the
${\cal WH}$-equation from 4 dimensions to the 2 dimensional case, we were
able to find simple evolution equations and also very simple
solutions for they.

The organization of this paper is as follows. In section
2 we find the
differential equation of "weak ${\cal H}$"-spaces using the basic spinorial
formalism. This derivation is based completely on an unpublished work by one
of us [7]. We show how the first Heavenly equation results in this
context and how the ${\cal WH}$-equation is a generalization of the
second Heavenly equation. The section 3 is devoted to perform a dimensional
reduction of
the ${\cal WH}$-equation considering only the two dimensional
case. We
write the ${\cal WH}$-equation in the Cauchy-Kovalevski form and we find
some
solutions by direct integration. Finally, in section 4, we give our
concluding remarks.

\vskip2cm

\centerline{ \bf 2. Differential Equations of Weak ${\cal H}$-Spaces}

\vskip 1truecm

\noindent
{\it (i).- Generalities}

In this section we start from the usual spinorial formulation for
the 2-form formalism for SD gravity on a four dimensional complex Riemannian
manifold ${\cal M}$ [8,9]. We would like to integrate the  equations:

$$d S^{\dot A\dot B} \ + \ 2 \alpha \ \wedge \ S^{\dot A\dot B} \ = \ 0
\eqno(2.1)$$
for $S^{\dot A\dot B}$ the anti-self-dual 2-form on ${\cal M}$ given by

$$ S^{\dot A\dot B} \ = \ {1\over 2} \ \epsilon_{RS} \ \ g^{R\dot A} \wedge
g^{S\dot B}, \eqno(2.2)$$
where $\dot A,\dot B \in \{\dot 1, \dot 2\}$, $\alpha$ is a 1-form on
 ${\cal M}$, $\epsilon_{RS}$ is the usual Levi-Civita's matrix
$(\epsilon_{RS}) =
\pmatrix{0 & 1 \cr -1 & 0} \ = \ (\epsilon^{RS})$ and $g^{A\dot B}$ is a
1-form on ${\cal M}$ which is the spinorial tetrad defining the Riemannian
metric as

$$ g = - {1\over 2} \ g_{A\dot B} \otimes g^{A\dot B}. \eqno(2.3)$$
here $\otimes$ denotes the tensor product.

At this level we can invoke one aspect of the Frobenius theorem which
implies
the existence of the scalar spinorial functions on ${\cal M}$

$$ \lambda^A_{\ B}, \ \tilde\lambda^A_{\ B}, \ q^A, \ \tilde q^A\eqno(2.4a)$$
being $\lambda^A_{\ B}$ and $\tilde \lambda^A_{\ B}$ constrained by

$$ {\rm det}(\lambda^A_{\ B}), \  {\rm det}(\tilde \lambda^A_{\ B}) \
\not= \ 0. \eqno(2.4b)$$

Moreover, we can put the spinorial tetrad as

$$ g^{A\dot 1} \ = \ \phi^{-1} \ e^\sigma \ dq^A; \ \ g^{A\dot 2} \ = \
\phi^{-1} \ e^{-\sigma} \ d \tilde q^A \eqno(2.5)$$
where we have used a convenient parametrization and the fact that this
tetrad
is obviously meaningful only modulo the SL(2,{\bf C})
transformations.\footnote{1}{ Notice
that the parametrization can be written as $g^{A\dot 1} = \phi^{-1}
e^{\sigma}
l^A_{\ B} dq^B$ and $g^{A\dot 2} = \phi^{-1} e^{-\sigma} {\tilde l}^A_{\
B}d\tilde q^B$. We can
take $g^{A\dot B} \rightarrow g^{\prime A\dot B} = L^A_{\ R} g^{R\dot B}$
with $(L^A_{\ B}) \in {\rm SL} (2, {\bf C})$
and det $(L^A_{\ B}) = 1$. Moreover, taking $L^A_{\ B} = - l^A_{\ B}$ we
have $g^{\prime A \dot 1}
= \phi^{-1} e^{\sigma} dq^A$, similarly for $g^{A \dot 2}$ we take $L^A_{\
B} = - \tilde {l}_B^{\ A}$.}

It is easy to see that substituting Eqs. (2.5) into (2.1) we obtain for
$\dot A \dot B \ = \ \dot 1 \dot 1$ and $\dot 2 \dot 2$, respectively

$$ (\alpha - d \ l n \ \phi \ + \ d \sigma) \wedge d q^1 \wedge d q^2 \ =
\ 0\eqno(2.6a)$$
and
$$ (\alpha - d \ l n \ \phi \ - \ d \sigma) \wedge d \tilde q^1 \wedge d \tilde
q^2
\ = \  0. \eqno(2.6b)$$

The pair $\big\{ q^A, \ \tilde q^A\big\}$ constitute a local chart on
${\cal M}$. So, Eqs. $(2.6a,b)$ and the equation for the volume form
$\omega =
-\phi^{-4} dq^1\wedge dq^2\wedge d\tilde q^1\wedge d\tilde q^2$ imply that
$\alpha$ can be written as

$$\alpha = d \ l n \ \phi \ + \ {\partial \sigma \over \partial q^A} \ d q^A \
- {\partial \sigma \over \partial \tilde q^A} \ d \tilde q^A. \eqno(2.7)$$

Taking the gauge on the spinorial tetrad (2.5) to be

$$g^{A\dot 1} \ = \ \phi^{-1} \ e^\sigma \ l^A_{\ B} \ d q^B, \ \
g^{A\dot 2}
= \phi^{-1} \ e^{-\sigma} \ \tilde l^A_{\ B} \ d \tilde q^B \eqno(2.8)$$
with det$(l^A_{\ B}) = 1 =$ det$(\tilde l^A_{\ B})$, it is possible to
express  the metric (2.3) in the form

$$ g = - \phi^{-2} \ \Omega_{AB} \ d q^A \ \otimes d \tilde q^B \eqno(2.9)$$
where $\Omega_{AB}: = \epsilon_{RS} \ l^R_{\ A} \ \ \tilde l^S_{\ B}$. This
definition satisfies
$${\rm det} \ (\Omega_{AB}) = {1\over 2} \ \Omega_{AB} \ \Omega^{AB} \ =
\ 1. \eqno(2.10)$$

Considering Eq. (2.1) with $\dot A \dot B = \dot 1 \dot 2$ the
integrability conditions are given by

$$ d \alpha \wedge S^{\dot 1 \dot 2} \ = \ 0, \eqno(2.11)$$
it can be reduced to the formula:

$$ \Omega^{AB} \ \ {\partial^2 \sigma \over \partial q^A \partial \tilde q^B}
\ = \ 0.  \eqno(2.12)$$

On the other hand, Eq. (2.2), for $\dot A  \dot B \ = \ \dot 1 \dot 2$ it is
found to be equivalent to a pair of scalar conditions

$$ {\partial \over \partial q^A}  \ (e^{2\sigma} \Omega^A_{\ B}) =0, \ \
{\partial
\over \partial \tilde q^B} \ (e^{-2\sigma} \Omega^{\ B}_A) =0. \eqno(2.13)$$

Manipulating these equations we obtain the following set
of differential equations

$$ {\partial^2 \Omega^{AB} \over \partial q^A \partial \tilde q^B} \ + \
4 \ \Omega^{AB} \ {\partial \sigma \over \partial q^A} \ \ {\partial \sigma
\over \partial \tilde q^B} \ = \ 0\eqno(2.14a) $$

$$ \Omega^{AB} \ \ {\partial^2 \sigma \over \partial q^A \partial \tilde q^B}
\ = \ 0. \eqno(2.14b)$$

Therefore Eq. (2.12)  is implicitly contained in Eqs. (2.13) and does
not represent an independent condition.

All this is very close to the description of ${\cal H}$-spaces theory via
the $\Omega$ formalism [8,9]. To see this we make $\sigma =$ {\it
constant}, which implies that Eqs. (2.13) are equivalent
to the existence of a scalar holomorphic function $\Omega =
\Omega(q_A,\tilde q_A)$ of its arguments such that

$$ \Omega_{AB} \ = \ {\partial^2 \Omega \over \partial q^A \partial \tilde
q^B}.
\eqno(2.15)$$
The condition det $(\Omega_{AB}) = 1$ reduces directly to the first heavenly
equation and the metric becomes conformally equivalent to the ${\cal H}$-space.

\vskip 2truecm
\noindent
{\it (ii).-The Weak Heavenly Equations}

We shall now seek for the equivalent formulation of the structure described by
Eqs. (2.13) and (2.14) corresponding to the $\Theta$-formalism for  the
${\cal H}$-spaces [8,9]. For this purpose, we observe first that Eqs. (2.13)
are equivalent to the existence of functions $P_A, \tilde P_A$ such that
they satisfy

$$ e^{2\sigma} \Omega_{AB} = {\partial \over \partial q^A} \ \tilde P_B, \ \ \
e^{-2\sigma} \Omega_{AB} \ = \ {\partial \over \partial \tilde q^B} \ P_A.
\eqno(2.16)$$

 From Eq. (2.16) and the condition det$(\Omega_{AB}) =1$, we infer that

$$ {\rm det} \bigg({\partial \over \partial q^A} \ \tilde P_B \bigg) \ =
e^{4\sigma} \ \ {\rm and} \ \  {\rm det}  \bigg( {\partial \over \partial
\tilde q^B} P_A \bigg) \ = \ e^{-4\sigma}. \eqno(2.17)$$
This means that the Jacobians

$${\partial (\tilde P_1, \tilde  P_2) \over \partial (q^\prime, q^2)} \ \ \
{\rm and} \ \ \ {\partial (P_1, P_2) \over \partial (\tilde q^\prime, \tilde
q^2)}$$
are different from zero. Thus, in place of the local
chart
$\{ q_A, \tilde q_A\}$ we can, alternatively, employ the charts $\{q_A,P_A\}$
and $\{ \tilde q_A, \tilde P_A \}.$

For the first chart $\{q_A,P_A\}$ the spinorial tetrad is

$$ g^{A\dot 1} =\chi^{-1} d q^A \eqno(2.18a)$$
$$ g^{A\dot 2} =\chi^{-1} (dP^A - Q^A_{\ B} \ d q^B) \eqno(2.18b)$$
with $P_A = P_A (q_A, \tilde q_A), \ \ \chi^{-1} \ = \ \phi e^{-\sigma}$
and $Q^A_{\ B} = {\partial P^A \over \partial q^B}.$ For the second chart
$\{\tilde q_A, \tilde P_A\}$ we have

$$ g^{A \dot 1} = - \tilde \chi^{-1} (d \tilde P^A - \tilde Q^A_{\ B}
d\tilde q^B) \eqno(2.19a)$$

$$ g^{A\dot 2} = \tilde \chi^{-1} d \tilde
q^A \eqno(2.19b)$$
where  $\tilde P_A = \tilde P_A (q_A, \tilde q_A), \ \tilde \chi^{-1} = \phi
e^{\sigma}$ and $\tilde Q^A_{\ B} = {\partial \tilde P^A \over \partial
\tilde q^B}.$

In both cases the metric takes the form

$$ g = - \chi^{-2} d q^A \otimes \ (d P_A - Q_{AB} \ d q^B),\eqno(2.20a)$$
and

$$ g = - \tilde\chi^{-2} d \tilde q^A \otimes (d \tilde P_A - \tilde
Q_{AB} \ d \tilde q^B). \eqno(2.20b)$$
Thus, apart from the conformal factors the metric is
determined entirely by the symmetric functions $Q_{(AB)}$ or $\tilde
Q_{(AB)}$, respectively.

On the other hand, one can write Eqs. (2.1) and (2.2) for $\dot
A\dot B \ = \ \dot 1\dot 1$ in the context of the $\Theta$-formalism.
These equations leads to

$$ (\alpha - d \ l n \ \chi) \wedge d q^1 \wedge d q^2 \ = \ 0 \eqno(2.21)$$
which implies that $\alpha$ takes the form

$$\alpha \ = \ d \ l n \chi \ + \ \alpha_A \ d q^A. \eqno(2.22)$$

Working with $S^{\dot 1\dot 2}$ from (2.2) we obtain

$$ (2 \alpha_A \ + \ {\partial \over \partial P^A} \ Q^S_{\ S}) \ d p^A
\wedge d q^1 \wedge d q^2  \ = \ 0. \eqno(2.23)$$
{}From this equation, it can be observed that

$$ \alpha_A \ = \ - \ {1\over 2} \ \ {\partial \over \partial P^A} \
Q^S_{\ S}, \eqno(2.24)$$
and consequently, Eq. (2.22) amounts to

$$ \alpha \ = \ d \ l n \ \chi - \ {1\over 2} \ {\partial \over \partial
P^A} \ Q^S_{\ S} \ d q^A. \eqno(2.25)$$

After some manipulations of Eq. (2.1) for $\dot A\dot B \ = \ \dot 2
\dot 2$ and cancelling terms involving $d \ l n \ \chi$, the factor
$\chi^{-2}$ and considering the fact that ${\rm det}(Q^A_{\ B}) = 0$, it
can be expressed as

$$\eqalign{& - d Q_{AB} \ \wedge \ d P^A \wedge d q^B \ + \ d( det
(Q^A_{\ B})) \ \wedge \ d
q^1 \ \wedge \ d q^2\cr& - {\partial \over \partial P^c}
Q^S_{\ S} d q^c \wedge [ dP^1 \wedge d p^2 - Q_{AB} d P^A \wedge d q^B] = 0.
\cr} \eqno(2.26)$$
Or, equivalently

$$\bigg\{ {\partial Q^B_{\ A} \over \partial P^B} \ \ - \ \ {\partial \over
\partial P^A} \ Q^B_{\ B}\ \bigg\} d q^A \wedge d P^1 \wedge d P^2
$$

$$+ \bigg\{ {\partial \over \partial P^A} \ {\rm det} \ (Q^R_{\ S}) + Q^{\
R}_A {\partial \over \partial P^R} \ Q^S_{\ S} \ - \
{\partial Q_A^{\ B} \over \partial q^B} \bigg\} d P^A \wedge d q^1 \wedge
d q^2 \ = \ 0.\eqno(2.27) $$

After some manipulations we can conclude that Eqs. (2.1) are fulfilled by
$S^{\dot A \dot B}$ in the chart $\{q_A,P_A\}$ if and only if the
structural functions $Q_A ^{\ B}$ fulfill the differential conditions

$$ {\partial \over \partial P^B} \ Q^{\ B}_A \ = \ 0, \eqno(2.28a)$$

$$ Q^{\ B}_S \ {\partial \over \partial P^B} \ \ Q^{\ S}_A \ + \ {\partial
\over \partial q^B} \ \ Q^{\ B}_A \ = \ 0. \eqno(2.28b)$$
with de corresponding 1-form $\alpha$ given by Eq. (2.25).

A similar procedure can be exactly realized with the other
equivalent chart $\{ \tilde q_A, \tilde P_A\}$  leading to similar
set of equations.

A direct consequence of Eqs. $(2.28a,b)$ is the existence of the scalar
holomorphic function of its arguments $\theta_A= \theta_A(q_A,P_A)$, which
satisfies

$$ Q_{AB} \ = \ {\partial \over \partial P^B} \ \theta_A.  \eqno(2.29)$$
At this level it is convenient to introduce the notational conventions

$$ \partial_A: \ = \ {\partial \over \partial P^A} , \ \ \ \ \ \ \ \
\partial^A: \ = \ {\partial \over \partial P_A}, $$

$$ \not \partial_A: \ = \ {\partial \over \partial q^A} , \ \ \ \ \ \ \ \
\not \partial^A: \ = \ {\partial \over \partial q_A}.  \eqno(2.30)$$
So, it is possible to give a more concise form for Eqs. (2.29) and
$(2.28b)$, they are respectively

$$ Q^B_{\ A} \ = \ - \partial^B \ \theta_A \eqno(2.31)$$
and

$$\partial^B \theta^R \cdot \partial_B \partial_R \theta_A \ - \ \partial^B
\not \partial_B \theta_A \ = \ 0. \eqno(2.32)$$

We shall consider they as the {\it fundamental equations} of the {\it weak
heavens} $(W {\cal H})$ in the $\theta$-formalism. Eq. (2.32) can be
expressed as

$$ \partial^B \ \{\partial_B \theta^R \cdot \partial_R \theta_A +
\not \partial_B \theta_A \} \ = \ 0, \eqno(2.33)$$
implying the ex\-is\-tence of a scalar func\-tion $\phi_A$ which
sat\-is\-fies  the con\-ser\-va\-tion law  $\partial^B \partial_B
\phi_A\- = \- 0$ with

$$ \partial_B \theta^R \cdot \partial_R \theta_A + \not \partial_B
\theta_A \ = \ \partial_B \phi_A. \eqno(2.34)$$

The structure of the ${\cal WH}$ (2.32) is intrinsically related to the
properties of the 1-form (2.25) which satisfies the proposition

$$ d \alpha \ = \ 0 \ \iff \ Q^S_{\ S} \ = - \partial^s \theta_s = {\partial
\over \partial q^A} {\cal R}(q)\cdot P^A + {\cal S},  \eqno(2.35)$$
where ${\cal R}$ and ${\cal S}$ are arbitrary
functions.

\vskip 2truecm
\noindent
{\it (iii).- Reduction to the First Heavenly Equation}

This case corresponds to the metrics conformally equivalent
to a right-flat space. To see this we consider the $\Omega$-formalism
[8,9]. According with Eq. (2.7) and (2.11), $\alpha$ can be expressed as

$$d\alpha = -2{\partial^2 \sigma \over \partial q^A \partial \tilde q^B}.
\eqno(2.36)$$
So, we have of course
$$ d\alpha = 0 \iff {\partial^2 \sigma \over \partial q^A \partial \tilde
q^B} = 0, \eqno(2.37)$$
which implies that $\sigma = \rho(q) + \tilde \rho(\tilde q)$. With
$\sigma$ of this form, we can equivalently spell out (2.13) in the form

$${\partial \Omega_{ \ B}^{\prime A} \over \partial q^A} = 0, \ \ \ \ \
{\partial \Omega_A^{\prime \ B}\over \partial \tilde q^B} = 0 \eqno(2.38)$$
where $\Omega'_{AB} = {\rm exp}(2(\rho - \tilde \rho)) \Omega_{AB}$.
This
equation implies the existence of the scalar functions $\Omega$ such that

$$\Omega'_{AB} = {\partial^2\Omega \over \partial q^A \partial \tilde
q^B} \Rightarrow \Omega = {\rm exp}(2(\tilde \rho - \rho))
{\partial^2\Omega \over \partial q^A \partial \tilde q^B}. \eqno(2.39)$$

The metric is now
$$g = - \phi^{-2} {\rm exp} (2(\rho - \tilde \rho)) {\partial^2 \Omega
\over \partial q^A \partial \tilde q^B} dq^A\otimes d\tilde q^B \eqno(2.40)$$
while $ {\rm det} ({\partial^2 \Omega \over \partial q^A \partial \tilde
q^B}) = 4 {\rm exp}(4(\rho - \tilde \rho))$.

In the last step, we execute the coordinate transformation in the
re-parametrization of the two congruences of the null strings

$$q^A= q^A(q'), \ \ \ \ \ \  \tilde q^A = \tilde q^A(\tilde q')\eqno(2.41)$$
with Jacobians so adjusted that

$${\partial(q')\over \partial(q)} = {\rm exp}(4\rho), \ \ \ \ {\partial
(\tilde q') \over \partial(\tilde q)} = {\rm exp}(-4 \tilde \rho).
\eqno(2.42)$$
They are different from zero. This reduces (2.40) to the first Heavenly
equation

$${\rm det} ({\partial^2 \Omega\over \partial q'^A \partial \tilde q'^B})
=1. \eqno(2.43)$$

The metric in the new chart $\{q'^A, \tilde q'^A\}$ is now

$$g = - \phi'^{-2} {\partial^2\Omega\over \partial q'^A \partial \tilde
q'^B} dq'^A \otimes d\tilde q'^B, \ \ \  \phi'= \phi {\rm exp} (\rho -
\tilde \rho). \eqno(2.44)$$

\vskip 2truecm

\noindent
{\it iv).- Generalization of the Second Heavenly Equation}

This SL$(2,{\bf C})$-gauge transformations leave the tetrad $(2.18a,b)$
form-invariant \footnote {2} {The SL$(2,{\bf C})$-gauge transformations
induced by $q^A = q^A(q')$, ${\partial (q')\over
\partial (q)} := {\rm exp} (4 \rho) \not=0 $, are ${\partial q^A\over
\partial q'^B} = {\rm exp}(-2 \rho) l^A_{\ B}$, ${\partial
q^{\prime A} \over
\partial q^B} = - {\rm exp}(2\rho) l^{\ A}_B$, ${\rm det}(l^A_{\ B}) =1$
where the matrix $(l^A_{\ B}) \in {\rm SL}(2,{\bf C})$.}

$$ g^{A\dot 1} \Rightarrow g'^{A\dot 1} = \chi'^{-1} dq'^A \eqno(2.45a)$$

$$ g^{A\dot 2} \Rightarrow g'^{A\dot 2} = \chi'^{-1}(dp'^A -
Q'^A_{\ B}dq'^B) \eqno(2.45b)$$
if and only if the structural functions $\chi$ and $Q^A_{\ B}$ transforms
as

$$\chi' = \chi {\rm exp}(2\rho)\eqno(2.46a)$$

$$P'^A = {\partial q'^A\over \partial q^B} P^B + \tau^A; \ \ \ {\partial
\tau^A\over \partial P^B} = 0 \eqno(2.46c)$$
and

$$Q'^A_{\ B} = {\partial q'^A\over \partial q^R} Q^R_{\ S}{\partial
q^S\over \partial q'^B} + P^R {\partial \over \partial q'^B}({\partial
q'^A \over \partial q^R}) + {\partial \tau^A \over \partial q'^B}.
\eqno(2.48)$$

Employing Eq. (2.25) in the primed version and Eq. (2.48) we can show
that assuming that Eqs. (2.1) are fulfilled, the 1-form $\alpha$ is
invariant under the discussed coordinate-tetrad transformation {\it i.e.}

$$\alpha ' = \alpha. \eqno(2.49)$$

Working in the $\{q_A,P_A\}$ chart, we take $d\alpha = 0$, then
according with (2.35) we have

$$Q^S_{\ S} = - {\partial{\cal R}(q) \over \partial q^S}\cdot P^S - {\cal
S}(q).\eqno(50a)$$
Passing to the chart $\{q'_A,P'_A\}$ and using (2.48) we arrive to

$$Q'^S_{\ S} = P^S {\partial \over \partial q^S} [ln {\partial(q')\over
\partial(q)} - {\cal R}(q)] + {\partial \tau^R\over \partial q'^R} -
{\cal S}(q). \eqno(2.50b)$$
Since ${\cal R}(q)$ and ${\cal S}(q)$ are arbitrary functions, without any
lost of generality we can take

$${\cal R}(q) = {\rm ln} {\partial (q')\over \partial (q)}, \ \ \ \ \  {\cal
S}(q) = {\partial \tau^R \over \partial q'^R}. \eqno(2.51)$$

Therefore assuming that $d\alpha = 0$ this implies that $Q^S_{\ S}$
vanishes, or

$$d \alpha = 0  \iff     Q^S_{\ S} = 0 \eqno(2.52)$$
and therefore from Eq. (2.31)

$$ Q^S_{\ S} \ = \ \partial^S \theta_S \ = \ 0. \eqno(2.53)$$
Thus, there exists the holomorphic scalar function $\Theta$ satisfying

$$ \theta_A \ = \ \partial_A \Theta \ = \ {\partial \over \partial P^A}
\Theta. \eqno(2.54)$$

Correspondingly

$$ Q_{AB} \ = \ \partial_A \partial_B \Theta \eqno(2.55)$$
is automatically symmetric. Substituting Eq. (2.55) in Eq. $(2.28b)$
we get that the $W{\cal H}$-equation reduces to

$$ -\partial^S \ \partial^R \ \Theta \ \cdot \ \partial_A \ \
(\partial_S \partial_R \Theta) \ \
- \ \ \partial^R \ \not \partial_R \ \partial_A \ \Theta \ = \ 0
\eqno(2.56)$$
or
$$ -\partial_A \ \bigg\{\ {1\over 2} \ \partial^S \partial^R \ \Theta \
\cdot \ \partial_S \
\partial_R \Theta \ + \ \partial^R \ \not \partial_R \ \Theta \ \bigg\} = 0
\eqno(2.57)$$
and so the $W{\cal H}$-equation reduces to the scalar condition

$$ {1\over 2} \ \ \partial^A \ \partial^B \ \Theta \ \cdot \ \partial_A \
\partial_B \ \Theta \
+ \ \partial^A \ \not \partial_A \ \Theta \ = \ \Xi (q). \eqno(2.58)$$

But with \ $Q_{AB} \ = \ \partial_A \partial_B \Theta$, \ we can send
$\Theta
\Rightarrow \Theta + \chi^A (q) P_A$, where $\chi^A (q)$ is chosen to be
a solution to $\not \partial_A \chi^A \ = \Xi (q)$. This maintains the basic
$Q_{AB} \ = \ \partial_A \partial_B \Theta$, reducing (2.58) to the
second heavenly equation in its standard form [8,9]

$$ {1\over 2} \ \partial^A \ \partial^B \ \Theta \ \cdot \ \partial_A \
\partial_B \ \Theta \
+ \ \partial^A \not \partial_B \Theta \ = \ 0. \eqno(2.57)$$

The metric is now

$$ g \ = \ -\chi^{-2} d q^A \otimes (d P_A - \partial_A \partial_B \Theta
d q^B) \eqno(2.60)$$
being thus conformally equivalent to the (strong) ${\cal H}$-spaces [8,9].

The basic point of this argument is that the $W{\cal H}$-equation
constitute a legitimate generalization of the second heavenly equation,
containing it as a special case.

Detailed computations about connections, curvature, Bianchi identities,
etc. can be found in Ref. [7].

\vskip2cm
\centerline{ \bf 3. The Evolution Weak Heavenly Equations}

In this section we first consider the ${\cal WH}$-equation given by
(2.28). After performing the dimensional reduction of this equation, we have

$$ {\partial \Theta \over \partial p} \ \cdot \ {\partial^2 \Theta \over
\partial
p^2} \ + \ {\partial^2 \Theta \over \partial p \partial q} \ = \ 0 \eqno(3.1)$$
defined on a two-dimensional manifold $\Sigma$ with coordinates
$\{ p, q\}$ and
where $p \ = \ P^1, \ \ q \ = \ q^1$ and $\theta^1 = \theta_1 =
\theta(p, q).$

Under this dimensional reduction the metric is

$$ g \ = \ -\chi^{-2} d q \otimes \ dp, \eqno(3.2)$$
the conformally flat metric, because $Q^1_{\ 1} = {\partial p \over
\partial q} = 0.$

For our purpose is more convenient to write (3.1) in the form
$$ \theta_{,p} \cdot \theta_{,pp} \ - \ \theta_{,pq} \ = \ 0. \eqno(3.3)$$
where $\theta_{,pq} \equiv {\partial^2 \theta \over \partial p \partial q}$.

In terms of differential forms the above equation is as follows

$$ d \theta - \theta_{,p} dp - \theta_{,q} d q \ = \ 0, \eqno(3.4 a)$$

$${1\over 2} d (\theta^2_{,p}) \wedge d q - d \theta_{,p} \wedge d p = 0.
\eqno(3.4 b)$$
Writing $(3.4 a)$ as

$$ d (\theta - p \theta_{,p}) + p d \theta_{,p} - \theta_{,q} d q = 0.
\eqno(3.5)$$
Now, performing the following Legendre transformation in the similar
spirit of [1]

$$ t = - \theta_{,p} \ \ \Rightarrow  \ \ \ p = p (t, q)$$

$$ {\cal H} = {\cal H} (t, q) = \theta (p (t, q), q) + t \cdot p (t, q),
\eqno(3.6)$$
from $(3.4 b)$ and (3.5) we easily find

$$ d {\cal H} - {\cal H}_{,t} d t - {\cal H}_{,q} d q \ = \ 0, \eqno(3.7 a)$$

$$ t d t \wedge d q + {\cal H}_{,tq} d t \wedge d q \ = \ 0. \eqno(3.7 b)$$

Thus, the above equations lead to

$${\cal H}_{,tq} = - t. \eqno(3.8)$$
a second order partial differential equation for a holomorphic function
${\cal H}(q,t)$ of its arguments. This is a very simple equation which
can be directly integrated out

$$ {\cal H} (t, q) = - {1\over 2} q t^2 \ + \ \int f (q) d q \eqno(3.9)$$
where $f(q)$ is a function which only depends on $q$.

Alternatively, we can choose the other coordinate in (3.4) to perform the
Legendre transformation

$$ t = - \theta_{,q} \ \ \ \ \ \Rightarrow \ \ \ \ \ q=q (t, p)$$

$${\cal H} = {\cal H} (t, p) = \theta (p, q, (t, p)) + t \cdot q (t, p).
\eqno(3.10)$$

Using $(3.4ab)$ and

$$ d (\theta - q \theta_{,q}) + q d \theta_{,q} - \theta_{,p} dp = 0
\eqno(3.11)$$
we find (taking $p \equiv x$)

$$ d {\cal H} - {\cal H}_{,t} d t - {\cal H}_{,x} dx = 0, \eqno(3.12 a)$$

$${1\over 2} d ({\cal H}^2_{,x}) \wedge d {\cal H}_{,t} - d {\cal H}_{,x}
\wedge dx = 0. \eqno(3.12 b)$$
Equivalently we obtain

$$ {\cal H}_{,xt} \ [ {\cal H}_{,xt} \cdot {\cal H}_{,x} - {\cal H}_{,xx}
\cdot {{\cal H}_{,tt} \over {\cal H}_{,xt}} + 1] = 0. \eqno(3.13)$$
Assuming that ${\cal H}_{,xt}$ is different from zero we find
that (3.13) reduces to

$$ {\cal H}_{,xx} {\cal H}_{,tt} - {\cal H}^2_{,xt} - (log {\cal
H}_{,x})_{,t} = 0. \eqno(3.14)$$
This equation appears to be the second heavenly equation for the two
variables $\{x,t\}$ with an additional logarithmic term.

It is possible to find formal solutions to (3.14) in the
sense
of  [1, 6], but instead of this, we would like to perform a second Legendre
transformation on it. This as an attempt to put this equation in a more
simple appearance and so, try to find simple solutions for it.

In terms of differential forms (3.14) gives

$$ d {\cal H} - {\cal H}_{,x} d x - {\cal H}_{,t} d t = 0, \eqno(3.15 a)$$

$$  d {\cal H}_{,x} \wedge d {\cal H}_{,t} + d (log {\cal H}_{,x}) \wedge
d x = 0. \eqno(3.15 b)$$

Performing the Legendre transformation

 $$ z = - {\cal H}_{,x} \ \ \ \ \ \Rightarrow \ \ \ \ \ x = x (z, t)$$

 $$ {\cal F} = {\cal F} (z, t) = {\cal H} ( x (z, t), t) + z \cdot
 x (z, t) \eqno(3.16)$$
{}From $(3.15 b)$ and

$$ d ({\cal H} - x {\cal H}_{,xx}) \ + \ x d {\cal H}_{,x} - {\cal H}_{,t}
\ dt \ = \ 0 \eqno(3.17)$$
one gets

$$ d {\cal F} \ - \ {\cal F}_{,z} d z \ - \ \ {\cal F}_{,t} \ d t \ = \ 0
\eqno(3.18 a)$$

$$ d (-z) \ \wedge d {\cal F}_{,t} \ + \  d (log (-z)) \wedge d {\cal
F}_{,z} \ = \ 0.  \eqno(3.18 b)$$
The logarithmic term imposes a constriction on the possible values of the
coordinate $z$. So, restricting our procedure to $z<0$ we find

$$ {\cal F}_{,tt} \ + \ {1\over z} \ {\cal F} _{,zt} \ = \ 0.   \eqno(3.19)$$

This equation has very simple solutions given by

$$ {\cal F} (z,t) \ = \ {\rm exp} \big[ k \ ({z^2 \over 2} + t)\big]
\eqno(3.20)$$
where $k$ is a constant.

It is remarkable that the Cauchy-Kovalevski form of the $W{\cal H}$-equation,
coincides with a modification of the second heavenly equation (3.14).

Also it is a very remarkable that the simplicity found in the
solutions for the ${\cal WH}$-equation in both cases.

\vskip 2truecm
\centerline{ \bf 4. Concluding Remarks}

In this paper we found a non-linear partial differential equation of the
second order that we have call {\it Weak Heavenly
equation} ${\cal WH}$. We showed that this equation is connected with
the first and second Heavenly equations in a close way. In fact, the ${\cal
WH}$
equation appears as a natural generalization for the second Heavenly
equation. It is claimed that the ${\cal WH}$-equation is a more fundamental
equations of SD
gravity than the Heavenly ones. We perform the dimensional reduction for the
${\cal WH}$-equation defined from the four dimensional manifold ${\cal M}$
to the two-manifold $\Sigma$. On $\Sigma$ we proved that the metric is
conformally flat. After the dimensional reduction we perform a Legendre
transformation on the coordinate $p$ and we found a very simple evolution
${\cal WH}$-equation which admits very simple solutions. The choice of
the coordinate $q$ to perform the Legendre transformation gave us a
differential equation which looks like the second Heavenly equation for
two dimensions. This
differential equation has an additional logarithmic term. Making a new
Legendre transformation on it we finally found a very simple differential
equation. This equation also admits simple solutions.

\vskip 2truecm
\centerline{\bf Acknowledgements}
One of us (H-G-C) wishes to thank CONACyT and SNI by support. Also he is
grateful to the staff of the Department of Physics at CINVESTAV, M\'exico
D.F. for the warm hospitality during the last six years. We are indebted to
L. Morales for useful suggestions.

\vfill\eject
\centerline{ \bf References}

\item{[1]} J.D.Finley III, J. F. Pleba\'nski, M. Przanowski and H.
Garc\'\i a- Compe\'an, {\it Phys. Lett.} {\bf A 181} 435 (1993).

\item{[2]} J. F. Pleba\'nski, M. Przanowski and H. Garc\'\i a-Compe\'an, The
Heavenly Equation has the Weak Painlev\'e Property, Preprint
CINVESTAV-IPN 93/06, (1993).

\item{[3]} J. F. Pleba\'nski, M. Przanowski and H. Garc\'\i a-Compe\'an,
{}From
Self-Dual Yang-Mills Fields to Self-Dual Gravity, Preprint CINVESTAV-IPN
93/07, submitted to {\it Phys. Lett.} {\bf B}.

\item{[4]} J. F. Pleba\'nski and M. Przanowski, Evolution Hyperheavenly
Equations, Preprint CINVESTAV-IPN 93/09 (1993), submitted to {\it J. Math.
Phys.}

\item{[5]} L.E. Morales, J. F. Pleba\'nski and M. Przanowski,
Cauchy-Kovalevskaya
Form of the Hyperheavenly Equations with Cosmological Constant, to be
published in {\it Rev. Mex. Fis.} {\bf 40} (1994).

\item{[6]} J.D.E. Grant, {\it Phys. Rev.} {\bf D48} 2606 (1993).

\item{[7]}  J.F. Pleba\'nski, The Differential Equations of Weak $H$-Spaces,
unpublished notes CINVESTAV-IPN (1985).

\item{[8]} J. F. Pleba\'nski, {\it J. Math. Phys.} {\bf 16} 2395 (1975).

\item{[9]} C.P. Boyer, J. D. Finley III and  J. F. Pleba\'nski, Complex
General Relativity, ${\cal H}$ and ${\cal HH}$ Spaces, A Survey of the
Approach, in General
Relativity and Gravitation,  Einstein's memorial volume, Ed. A. Held,
Plenum Press, New York (1980) Vol. 2 pp 241-281.

\bye